\documentclass[twocolumn]{aastex63}
\usepackage[latin1]{inputenc}
\usepackage{amsmath}
\usepackage{amsfonts}
\usepackage{amssymb}
\usepackage{subfigure}
\usepackage{graphicx}
\usepackage{xcolor}

\usepackage{natbib}
\newcommand{\steady}[0]{\textit{16TI} }
\newcommand{\spikes}[0]{\textit{40sp\_down} }

\listofchanges
\begin{document}

\title{Photospheric Polarization Signatures From Long Gamma Ray Burst Simulations} 
\author{Tyler Parsotan$^1$, Diego L{\'o}pez-C{\'a}mara$^2$, and Davide Lazzati}
\affiliation{Department of Physics, Oregon State University, 301 Weniger Hall, Corvallis, OR 97331, U.S.A.\\$^2$CONACyT - Instituto de Astronom{\'i}a, Universidad Nacional Aut{\'o}noma de M{\'e}xico, A. P. 70-264 04510 CDMX, Mexico}

\begin{abstract} 
A comprehensive understanding of Gamma Ray Bursts (GRBs) has been elusive due to the variety of questions surrounding the radiation mechanism at play in these events. Polarization measurements of GRBs can heavily constrain the relevant radiation mechanisms and the structure of the GRB jet; however, there is a limited number of theoretical predictions that observed GRB polarizations can be compared against. Here, we conduct radiative transfer calculations of a set of two dimensional relativistic hydrodynamic long GRB (LGRB) jet simulations, of a constant and a variable jet, using the Monte Carlo Radiation Transport (MCRaT) code. MCRaT has been enhanced by the inclusion of polarization; it has been first verified by reproducing a variety of results in the literature and then used to obtain the time integrated and resolved polarization degrees and angles of the synthetic LGRBs. While the obtained time-integrated polarization degrees {($\lesssim 1$\%)} are consistent with the constraints from the POLAR experiment, they are lower than other theoretical studies due to the lack of strong gradients in the model jet profiles that we use. {The time resolved results suggests that GRBs with wide jets observed on axis will have small polarization degrees ($\lesssim 2\%$) and \added{constant polarization angles}, during the brightest portion of the light curve. GRBs observed off axis will have larger polarization degrees and polarization angles that change with the temporal structure of radiating shells in the outflow. We then place our results in the context of GRB prompt emission models and future LEAP and POLAR-2 GRB polarimetry detections.} \newline
\end{abstract} 

%\maketitle

\section{Introduction}
Gamma Ray Bursts have been detected since the late 1960's as a relatively short pulse of gamma ray photons \citep{first_grbs}. These transient events have been categorized based on their duration. Events that last $\lesssim 2$ seconds are Short GRBs and are associated with the merger of compact objects \citep{GW_NS_merger, grb_NS_merger_connection, lazzati2018_GRB170817_afterglow} while events that last for $\gtrsim 2$ seconds are denoted Long GRBs (LGRBs) and are associated with core collapse supernovae \citep{grb_sn_connection, grb_collapsar_model, hjorth2003_LGRB_SNe}. Regardless of the type of GRB that is observed, the physical mechanism that produces the high energy X-ray and $\gamma$-ray photons that are observed during the first few seconds of these events, known as the prompt emission, is still under investigation.

There are currently two major competing theories: the synchrotron model (SM) \citep{SSM_REES_MES, ICMART_Zhang_2010} and the photospheric model \citep{REES_MES_dissipative_photosphere, Peer_photospheric_non-thermal,Belo_collisional_photospheric_heating, lazzati_variable_photosphere}. The {optically thin} SM describes shells of material, which have been launched with varying speeds by a central engine, colliding with one another far from the central engine. These collisions produce non-thermal radiation which is able to escape the jet if the opacity $\tau < 1$. This model is able to account for general characteristics of GRBs such as variability and non-thermal spectra, but is in tension with observational relationships such as the Amati, Yonetoku and Golenetskii Correlations {\citep{Amati,Yonetoku,Golenetskii, ICMART_Zhang_2010}. \added{Although \cite{mochkovitch2015_amati_synch} showed that the internal shock model can satisfy the Amati relation under certain conditions, there have been other subcategories of synchrotron models developed in an attempt to rectify these discrepancies.} These models consider the effects of both globally ordered and random magnetic fields \citep{toma2008statistical_GRB_pol, ICMART_Zhang_2010}.}

On the other hand, the photospheric model follows photons that have been produced deep in the jet. These photons interact with the matter in the jet until the jet becomes transparent to the radiation. Unlike the SM, this model is able to reproduce most of the observational relationships \citep{lazzati_photopshere, diego_lazzati_variable_grb}. Subphotospheric dissipation events \citep{Atul} and the idea of the photospheric region, in which the photosphere is a volume of space in which photons can still be upscattered by sparse interactions with matter in the jet \citep{parsotan_mcrat, parsotan_var, Ito_3D_RHD, Peer_fuzzy_photosphere, Beloborodov_fuzzy_photosphere, ito2019photospheric}, contribute to the non-thermal nature of the spectra in the photospheric model. %add mcrat_vary paper here
Although this model is able to reproduce general characteristics of GRBs, it is not able to fully account for the relatively large amount of low energy photons that are observed in GRB spectra.

%talk about observed GRB polarizations eg POLAR, etc.
The shortcomings of each model have placed them on relatively equal footing; however, polarization measurements of the prompt emission can help break this degeneracy. While there have been a number of polarization measurements made by a variety of instruments, the results have been largely inconclusive due to the difficulty of conducting such an observation with high precision.  From the past decade, the largest {linear} polarization measurement reported was $98 \pm 33$\% from GRB 041219A \citep{kalemci2007_high_pol} and the smallest was $27 \pm 11$\%  from GRB 100826A in a time resolved analysis \citep{yonetoku2011_min_pol}, where the reported errors are $1\sigma$ (also see \cite{Gill_Polarization} for a comprehensive list of detected GRB polarizations measurements). These measurements and many others are time integrated in order to get as much signal as is possible, however the uncertainties are still very large. These relatively high polarizations are typically interpreted under the assumption that only synchrotron radiation can produce such high polarizations \citep{waxman2003_synch_pol, lyutikov2003_synch_pol,burgess2019polar}. \added{A number of studies have shown that GRB jets with ordered magnetic fields can produce  high polarizations ranging between $20\%$ and $70\%$ (see e.g. \cite{deng2016collision},  \cite{lan2020GRB_time_pol}, \cite{toma2008statistical_GRB_pol} and \cite{Gill_Polarization})}, while jets with random magnetic fields produce smaller polarizations. Photospheric emission was originally thought to only produce very small polarizations, however, it has been shown that this model can produce polarization up to $\sim 40$\% if the jet has significant structure within $\delta \theta \sim \Gamma^{-1}$, where $\Gamma$ is the bulk Lorentz factor of the jet \citep{Lundman_polarization, lundman2018polarization, ito_polarization}, although this configuration may not occur in GRBs. Structure in the jet refers to gradients in the jet profile and/or anisotropy in the outflow of photons (which is related to the expanding outflow) \citep{lundman2014polarization}. If the source of soft photons in the photospheric model is due to synchrotron emission then the detected polarization can increase up to  $\sim 50$\% \citep{lundman2018polarization}.

Building on the polarimetry technology used to acquire past measurements, the POLAR experiment \citep{polar_detector} recently reported time integrated linear polarizations for 5 GRBs and had enough statistics to conduct a time resolved polarization analysis for one of the GRBs, GRB 170114A \citep{zhang2019polar}. Their analysis showed that the GRBs had relatively small upper limits for their time integrated linear polarizations, with the largest upper limit being $68$\% and the smallest upper limit being $28$\%; typically they find that the linear polarizations are $\lesssim 10$\%. {The authors claim that although GRB 170114A had a small time integrated linear polarization, the time resolved portions of the GRB show relatively high polarization ($\gtrsim 10$\%) with a continually changing polarization angle.} In acquiring these results, {\cite{zhang2019polar} make a number of assumptions about the physics of the GRB polarization such as assuming that polarization degree is constant throughout a GRB and that polarization angle can change in time. }

The authors interpret the changing polarization angle to be in strain with the photospheric model. Following this analysis, \cite{burgess2019polar} analyzed GRB 170114A by combining information from Fermi, which also observed the GRB. {They find similar results in that the SSM seems to describe the GRB well despite the other weakness associated with the SSM model. There is ambiguity, however, since the low linear polarization degree of GRB 170114A is easily attainable in the photospheric model.} As a result, \cite{burgess2019polar} call for more theoretical modeling and predictions
%, of the SSM and the photospheric model, 
of time resolved polarization signatures that will inform future analysis.

There have been no prior analyses of time integrated or time resolved polarization angle or polarization degree under the photospheric model using the realistic profile of a GRB jet. In this work we focus on the photospheric model and show that it is able to account for a changing polarization angle and variable polarization degrees. We use the MCRaT (Monte Carlo Radiation Transfer) code {\footnote{The MCRaT code is open-source and is available to download at: https://github.com/lazzati-astro/MCRaT/}} to provide time integrated polarization predictions and the first time resolved analysis of {LGRB} simulations using the photospheric model. In Section \ref{global_methods} we describe the methods we use to conduct mock observations of our synthetic LGRBs and outline how polarization is handled in MCRaT. Finally, in Section \ref{results} and Section \ref{summary}, we present the results and discuss the implications to GRB polarimetry missions and understanding the radiation mechanism in GRBs.

\section{Methods}
\label{global_methods}
There are numerous works that have explored polarization in the energy regime applicable to GRBs. These works have focused primarily on the Stokes parameters formalism (see eg \cite{McMaster:1961aa, ito_polarization, lundman2014polarization, depaola2003new, krawczynski2011polarization}). Here, we describe how we conduct mock observations of polarization degree and polarization angle, and then we describe the implementation of polarization in MCRaT and show that it is able to reproduce the results of \cite{depaola2003new}, \cite{krawczynski2011polarization}, and \cite{lundman2014polarization}. For more in depth discussions of the Stokes parameters we refer the reader to the aforementioned references. Finally, we outline how we determine equal time of arrival surfaces within the hydrodynamic simulations used in this work. 

\subsection{Mock Observations of Polarization}
\label{methods}
We produce light curves and spectra in the same manner outlined in \cite{parsotan_mcrat} and \cite{parsotan_var}, by collecting photons within a given viewing angle and binning them in time and energy. Here, we outline how we calculate the detected polarization degree, the polarization angle and their respective errors.

The Stokes parameters are a vector, $S=(I, Q, U, V)$ that holds information about the polarization of electromagnetic radiation. $I$ is the intensity of the electromagnetic radiation, $Q$ and $U$ describe the orientation of the polarization ellipse, and $V$ describes the ratio of the principal axis of the polarization ellipse \citep{Rybiki_Lightman}. {We follow the convention set by \cite{McMaster:1961aa} and  \cite{lundman2014polarization}
where $Q=+1$ is oriented with the y-axis of the Stokes plane and $Q=-1$ is oriented with the x-axis of the Stokes plane. The $+U$ axis is rotated $45^\circ$ clockwise with respect to the $+Q$ axis and the  $-U$ axis is rotated $45^\circ$ clockwise with respect to the $-Q$ axis. }Furthermore, we normalize the Stokes parameters such that $I=1$ at all times giving us $s=(1, Q/I, U/I, V/I)=(1,q,u,v)$. In our simulations, we only consider linear polarization which means that we ignore any contribution by $v$. This is appropriate since we assume that electron spins, which directly affect $v$, are isotropically distributed.

The polarization degree, $\Pi$, represents the average polarization of the detected photons \citep{Rybiki_Lightman}. From the stokes parameters $\Pi$ is calculated as:
\begin{equation}
\Pi=\sqrt{\bigg( \frac{Q}{I} \bigg)^2 + \bigg( \frac{U}{I} \bigg)^2 + \bigg( \frac{V}{I} \bigg)^2 }=\sqrt{q^2+u^2}
\end{equation}
where the second portion of the equation takes the normalization by $I$ into account and ignores $v$ since we do not consider circular polarization. Since the photons in MCRaT are weighted to increase computational efficiency (one photon packet in MCRaT represents some number of real photons in the relativistic outflow), we cannot simply add each photon's detected Stokes parameters to calculate $q$ and $u$; instead, we have to take the photons weight, $w$, into consideration by averaging the detected photons' Stokes parameters (see \cite{parsotan_var} for a discussion of the weight). Thus, we calculate $q$ and $u$ as:
\begin{equation}
q=\frac{\sum  w_iq_i}{\sum w_i} \qquad
u=\frac{\sum w_iu_i}{\sum w_i}
\end{equation}

The error in the polarization degree, $\sigma_\Pi$, is given by \cite{kislat2015_pol_error} as:
\begin{equation}
\sigma_\Pi \approx \sqrt{\frac{2-\Pi^2\mu^2}{(N-1)\mu^2}}
\end{equation}
where $N$ is the number of photons that were detected and $\mu$ is the modulation factor. For a perfect detector $\mu=1$, which is what we assume in this work. Additionally, the formulas acquired by \cite{kislat2015_pol_error}, that we use in this work, slightly underestimate the MCRaT error bars since they do not consider variances in photon weights. Following the analysis presented in their appendix, we multiply each error bar by $\sqrt{<w^2>/<w>^2}$ to account for this factor, where the angles brackets denote averages.

{The polarization angle, $\chi$, represents the net direction of the electric field vector once all the detected photons have been summed over \citep{kislat2015_pol_error}. $\chi$, the angle between the $+q$ axis and the electric field vector, measured clockwise towards the $+u$ axis of the stokes plane, is given by \cite{kislat2015_pol_error} as}:
\begin{equation}
\chi=\frac{1}{2}\arctan \big(\frac{u}{q} \big)
\end{equation}
The error in the polarization angle, as given by \cite{kislat2015_pol_error}, is:
\begin{equation}
\sigma_\chi\approx \frac{1}{\Pi \mu \sqrt{2(N-1)}}
\end{equation}
For the simulations analyzed in this work, where we assume an axisymmetric geometry, $\chi$ should be aligned with the positive or negative Stokes $Q$ values due to the sum of the $U$ parameters adding to zero \citep{lundman2014polarization}. We verified that the number of photons used in this work are high enough that $\sum u \approx 0$.  Additionally, $\chi$ has $\pi$ symmetry so we plot it between $-90^\circ$ and $+90^\circ$.

\subsection{Polarization in the MCRaT code} \label{mcrat}
%general discussion of polarization implementation and verification of implementation
In MCRaT all photons are initialized to have no polarization; thus, we set $s=(1,0,0,0)$ similar to \cite{lundman2014polarization}. Each photon becomes $100$\% polarized from the very first scattering that it undergoes, which does not bias our results. As mentioned before, we only consider linear polarization which means that we ignore any contribution by $v$. 

In order to deal with polarization, we follow the prescription given by \cite{lundman2014polarization}. We lorentz boost the photons' four momenta from the lab frame to the fluid frame, then from the fluid frame to the electron rest frame. In the electron frame we conduct our scattering using the full Klein Nishina (KN) cross section and then scatter the Stokes parameters using Fanno's Matrix. Afterwards, we deboost the photons from the electron rest frame and the fluid rest frame back to the lab frame. 

{In the lab frame, the Stokes plane is oriented such that the $+Q$ axis is pointing in the $-\hat{\phi}$ direction and the $-Q$ axis is pointing in the $-\hat{\theta}$ direction. The $+U$ axis is rotated $45^\circ$ clockwise with respect to the $+Q$ axis and the  $-U$ axis is rotated $45^\circ$ clockwise with respect to the $-Q$ axis. $\chi$ is measures clockwise from the $+Q$ axis towards the $+U$ axis. $\hat{\phi}$ and $\hat{\theta}$ are the orthonormal azimuthal and polar unit vectors in a spherical coordinate system where the radial unit vector is parallel to the photon's momentum vector.} This setup is a natural outcome of using \citeauthor{lundman2014polarization}'s (\citeyear{lundman2014polarization}) definitions.

Each boost to another reference frame entails rotating the Stokes plane using the Muller rotation matrix \citep{McMaster:1961aa} given by
\begin{equation}
\mathbf{M}[\phi]=\left(\begin{array}{cccc}{1} & {0} & {0} & {0} \\ {0} & {\cos 2 \phi} & {-\sin 2 \phi} & {0} \\ {0} & {\sin 2 \phi} & {\cos 2 \phi} & {0} \\ {0} & {0} & {0} & {1}\end{array}\right)
\end{equation}
where the angle of rotation, $\phi$, corresponds to the rotation that orients the y-axis of the Stokes plane perpendicular to the photon three momentum and the velocity vector of the frame that the photon will be boosted into. The equation for $\phi$ is given in Appendix B of \cite{lundman2014polarization}. After each boost, we rotate the Stokes plane again to ensure that the y-axis of the Stokes plane is perpendicular to the plane formed by the reference frame's z-axis and the photon's three momentum.

We use the KN cross section to determine if the photon scatters. The photons gets scattered if $\xi \le \sigma_{\mathrm{KN}}/\sigma_{\mathrm{T}}$, where $\xi$ is a random uniformly distributed number between 0 and 1, $\sigma_{\mathrm{KN}}$, is the KN cross section given by \cite{Rybiki_Lightman}, and $\sigma_{\mathrm{T}}$ is the Thomson cross section. If $\xi \le \sigma_{\mathrm{KN}}/\sigma_{\mathrm{T}}$ then we sample the differential KN cross section to determine the angles, $\theta_{\mathrm{sc}}$ and $\phi_{\mathrm{sc}}$, that the photon will be scattered into. The differential cross section given by \cite{lundman2014polarization} is
\begin{multline}
\frac{\mathrm{d} \sigma_{\mathrm{KN}}}{\mathrm{d} \Omega} (\theta_{\mathrm{sc}}, \phi_{\mathrm{sc}}) =\frac{r_{0}^{2}}{2}\left(\frac{\epsilon}{\epsilon_{0}}\right)^{2} \times \\
\left\{\frac{\epsilon_{0}}{\epsilon}+\frac{\epsilon}{\epsilon_{0}} - \sin ^{2} \theta_{\mathrm{sc}}\left(1- q \cos 2 \phi_{\mathrm{sc}}+u \sin 2 \phi_{\mathrm{sc}}\right)\right\} \label{KN_diff_cross_section}
\end{multline}
where $r_{0}$ is the classical electron radius, $\epsilon_{0}$ is the incoming photon energy scaled by the electron rest mass, and $\epsilon=\epsilon_{0}/(1+\epsilon_{0}[1-\cos \theta_{\mathrm{sc}}])$, is the scattered photons energy. The outgoing photon's $\theta_{\mathrm{sc}}$ is acquired by rejection sampling the differential cross section integrated over $\phi_{\mathrm{sc}}$ following the method outlined by \cite{sample_kn_theta} for maximum efficiency. We acquire $\phi_{\mathrm{sc}}$ by choosing a random uniformly distributed value between [0, 2$\pi$] when $q=u=0$, otherwise we apply a rejection sampling method to the differential KN cross section. {The case of $q=u=0$ removes the KN cross section's  dependence on $\phi_{\mathrm{sc}}$, which allows us to choose a value between [0, 2$\pi$].}

\cite{depaola2003new} presents a method of sampling the KN azimuthal distribution, however they use a KN cross section that is in a different form than the one that is employed above. Thus, we outline our method of sampling Equation \ref{KN_diff_cross_section} to acquire $\phi_{\mathrm{sc}}$ when the incoming photon is polarized. The azimuthal angle that gives the maximum of the KN differential cross section, $\phi_{\mathrm{m}}$, can be solved as
\begin{equation}
\phi_{\mathrm{m}}=\frac{1}{2} | \arctan \big( \frac{-u}{q} \big) |
\end{equation}
We then plug $\phi_{\mathrm{m}}$ and the previously acquired $\theta_{\mathrm{sc}}$ into the KN differential cross section to get the normalization for the rejection sampling method, $\frac{\mathrm{d} \sigma_{\mathrm{KN}}}{\mathrm{d} \Omega} (\theta_{\mathrm{sc}}, \phi_{\mathrm{m}})$. We proceed as follows:
\begin{itemize}
\item[(1)] Draw a random number $\xi$ uniformly distributed between 0 and 1
\item[(2)] Draw a random $\phi_{\mathrm{sc}}$ uniformly distributed between 0 and $2\pi$
\item[(3)] Calculate $\frac{\mathrm{d} \sigma_{\mathrm{KN}}}{\mathrm{d} \Omega} (\theta_{\mathrm{sc}}, \phi_{\mathrm{sc}})$
\item[(4)] Determine if $\xi \le  [\frac{\mathrm{d} \sigma_{\mathrm{KN}}}{\mathrm{d} \Omega} (\theta_{\mathrm{sc}}, \phi_{\mathrm{sc}})][\frac{\mathrm{d} \sigma_{\mathrm{KN}}}{\mathrm{d} \Omega} (\theta_{\mathrm{sc}}, \phi_{\mathrm{m}})]^{-1}$
\item[(5)] If the above condition is met, accept the value of $\phi_{\mathrm{sc}}$, otherwise repeat the sampling
\end{itemize}  

Once the scattering is completed, we conduct the scattering of the Stokes parameters. First, we use the Muller matrix to rotate the Stokes plane such that the Stokes plane's y-axis is perpendicular to the plane formed by the incoming and outgoing photon three momenta \citep{McMaster:1961aa, lundman2014polarization}. Then, we use Fanno's matrix, $T[\theta_{\mathrm{sc}},\epsilon_{0}, \epsilon]$, \citep{McMaster:1961aa} to determine the resulting polarization, $\tilde{s}=T[\theta_{\mathrm{sc}},\epsilon_{0}, \epsilon]s$. Fanno's matrix is 
\begin{multline}
T[\theta_{\mathrm{sc}},\epsilon_{0}, \epsilon]= \\ \left(\begin{array}{lll}{1+\cos ^{2} \theta_{\mathrm{sc}}+\left(\epsilon_{0}-\epsilon \right)\left(1-\cos \theta_{\mathrm{sc}}\right)} & {\sin ^{2} \theta_{\mathrm{sc}}} & {0} \\ {\sin ^{2} \theta_{\mathrm{sc}}} & {1+\cos ^{2} \theta_{\mathrm{sc}}} & {0} \\ {0} & {0} & {2 \cos \theta_{\mathrm{sc}}}\end{array}\right)
\end{multline}
where, similar to \cite{krawczynski2011polarization},  we have excluded the factors in front since they cancel out when we normalize the scattered Stokes parameters by $I$ as mentioned above. We have also excluded the fourth row and column of the matrix since we only consider linear polarization. 

%In order to verify the sampling algorithm of the differential KN cross section, we reproduce the results of \cite{depaola2003new}.
%we reproduce Figures, 3, 5 and 6 of \cite{depaola2003new}. 
%To test the Lorentz transform portion of MCRaT we reproduce the results presented by \cite{krawczynski2011polarization}.
%in particular their Figures 4 and 11, 
%Some of the MCRaT results reproducing \citeauthor{krawczynski2011polarization}'s  (\citeyear{krawczynski2011polarization}) Figure 6 and  \citeauthor{depaola2003new}'s  (\citeyear{depaola2003new}) Figure 3 are presented in Appendix \ref{comp_mcrat}. 
{We reproduce \citeauthor{depaola2003new}'s  (\citeyear{depaola2003new}) result, as shown in Figure \ref{depaola_comparison},  which verifies the sampling algorithm of the differential KN cross section. Additionally, reproducing \citeauthor{krawczynski2011polarization}'s  (\citeyear{krawczynski2011polarization}) results, in Figure \ref{kraw_comparison}, tests the Lorentz transform portion of MCRaT.}

In order to test the code globally, we reproduce the results of \cite{lundman2014polarization}. To do so, we imposed the analytic jet structure provided by \cite{lundman2014polarization} on the same simulation frames that we will use in Section \ref{results}. This is similar to \cite{MCRaT} simulating a variety of outflows by imposing an analytic solution onto hydrodynamic simulation files. {The domain of this simulation is $2.5 \times 10^{13}$ cm along the direction of the jet and $5 \times 10^{12}$ cm along the x axis.} We used $\sim 6 \times 10^5$ photons to conduct our code validation for a wide structured jet with $\theta_{\mathrm{j}}=0.1$ radians ($\sim 5.7^\circ$), $\Gamma_0=100$ and $L=3 \times 10^{50}$ erg/s. This is the same case exhibited in \citeauthor{lundman2014polarization}'s (\citeyear{lundman2014polarization}) wide jet with the exception of the value of $L$ that we use, which was chosen to maximize the number of photons that reached the photosphere before they approached the edge of the domain of the hydrodynamic simulation. {There are two major differences between the simulation conducted by \cite{lundman2014polarization} and the MCRaT simulation: 1) MCRaT uses the full Klein Nishina Cross section to determine if photons scatter while \citeauthor{lundman2014polarization}'s (\citeyear{lundman2014polarization}) simulation  uses the Thomson cross section and 2) The photons in MCRaT are not permitted to immediately escape to infinity if the randomly drawn optical depth is small enough while photons in  \citeauthor{lundman2014polarization}'s (\citeyear{lundman2014polarization}) simulation are allowed to do so. }

%\added{\citeauthor{lundman2014polarization} (\citeyear{lundman2014polarization}) state that they have a minimum of 200 photons in each angle bin, compared to the minimum number of photons used to produce the MCRaT data points in Figure \ref{fig:complundmanp4thetaj1} of $\sim 9 \times 10^3$, with an average number of photons in each angle bin of $\sim 3.8 \times 10^4 $.}

\begin{figure}
\centering
\includegraphics[width=\linewidth]{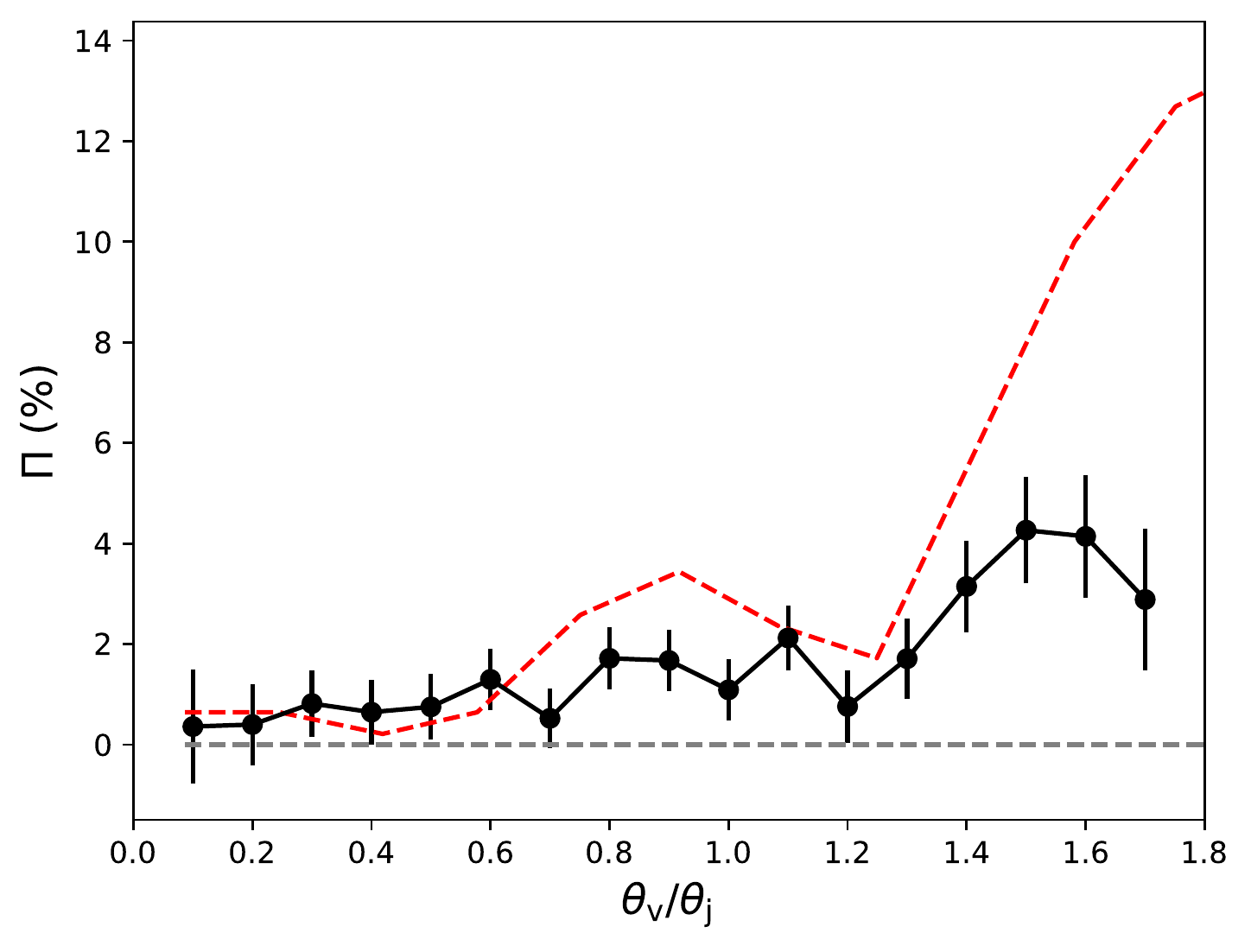}
\caption{A comparison between the polarization acquired by \cite{lundman2014polarization} for a structured jet with $\theta_{\mathrm{j}}=0.1$ radians $\Gamma_0=100$ and $L=3 \times 10^{50}$ erg/s, shown by the {red} dashed line, and the MCRaT acquired polarization shown by the black points with $1\sigma$ error bars. The dashed grey line denotes $\Pi=0\%$ for reference. We find agreement until  $\theta_{\mathrm{v}}/\theta_{\mathrm{j}} \approx 1.6$ ($\theta_{\mathrm{v}} \approx 9^\circ$) where the MCRaT photons are no longer decoupled from the flow by the time they reach the edge of the simulation domain. }
\label{fig:complundmanp4thetaj1}
\end{figure}

In Figure \ref{fig:complundmanp4thetaj1} we show the results of our validation. The polarization acquired by \cite{lundman2014polarization} is shown as the red dotted line, the black points with $1\sigma$ error bars show the polarization acquired by MCRaT, and the grey dotted line shows $\Pi=0\%$ for reference. In addition to finding that the Stokes $u$ parameter vanishes ($\sum u \approx 0$, see section \ref{methods}), which is expected for an axis symmetric jet \citep{ito_polarization, lundman2014polarization}, we find that MCRaT is able to recover \citeauthor{lundman2014polarization}'s (\citeyear{lundman2014polarization}) polarization profile relatively well. {We are also able to recover the change in the sign of the stokes Q parameter at $\theta_{\mathrm{v}}/\theta_{\mathrm{j}} \sim 1.2$ that \cite{lundman2014polarization} find in their results.
The MCRaT result is slightly lower than the polarization acquired by \cite{lundman2014polarization} due to the fact that the analytic jet profile is mapped onto a discretized grid; this has the effect of decreasing the gradients in the jet profile that would contribute to a larger polarization. 
Furthermore, \citeauthor{lundman2014polarization}'s (\citeyear{lundman2014polarization}) polarization profile contains a minimum of 200 photons in each angle bin, while the results of the MCRaT validation contain at least $\sim 10000$ photons in each angle bin, increasing the general precision of the MCRaT results. }

For $\theta_{\mathrm{v}}/\theta_{\mathrm{j}} \gtrsim 1.6$ ($\theta_{\mathrm{v}} \approx 9^\circ$) the MCRaT polarization begins to decrease again, coming into strain with what is expected. This is due to the fact that the simulation files that we impose the analytic jet equations onto have a finite domain. Even when the photons reach the edge of the domain ($2.5 \times 10^{13}$ cm) at $\theta_{\mathrm{v}} \gtrsim 9^\circ$, they haven't fully decoupled from the photosphere (which would be located at $r\sim 1\times 10^{14}$ cm \citep{lundman2014polarization}) thus decreasing the detected polarization. 

%As an additional check, we expect the Stokes $U$ parameter to vanish as they are added due to the symmetry of the jet about its axis \cite{lundman2014polarization}. We find that our Stokes $U$ parameters vanish with increasing number of photons, adding to the validity of the implementation.

\subsection{Equal Time of Arrival Surfaces}
\label{surfaces}
In order to relate the structure of the jet in the GRB simulations to the time dependent observables produced by MCRaT {(such as the time resolved polarization)}, we need to calculate the location  that photons would be emitting along the observer's line of sight for a given time in the light curve, $t_\text{detect}$. We follow the derivation for \citeauthor{parsotan_mcrat}'s (\citeyear{parsotan_mcrat}) Equation 1. The time in which a photon would be emitted from the jet, $t_j$, can be calculated as
\begin{equation}
    t_j=t_\text{detect}-t_\text{real}
\end{equation}
where $t_\text{real}$ is the real time acquired from the hydrodynamic simulation. The radius, $r_j$, at which a photon is emitted by the jet is
\begin{equation}
    r_j=t_j c
\end{equation}
where $c$ is the speed of light. The equal time of arrival surface {\citep{Zhang_E_p_evolution, Peer_multicolor_bb, beloborodov2011radiative}} is then constructed by drawing a line tangent to the circle that passes through $r_j$ and the observer viewing angle, $\theta_\mathrm{v}$ 
%{(see eg Figure \ref{time_surfaces}).}
 
\section{Results} \label{results}
In this work, we ran MCRaT on two different FLASH {2D} special relativistic hydrodynamic (RHD) simulations which both launched a jet into a 16TI progenitor star with a density profile provided by \cite{Woosley_Heger}. The first simulation which we call the \steady simulation, has a jet injection radius of $1\times 10^9$ cm, an initial lorentz factor of 5, an opening angle of $10^\circ$, an internal over rest-mass energy ratio, $\eta=80$, and the engine was active for 100 s \citep{lazzati_photopshere}. The domain of this simulation is $2.5 \times 10^{13}$ cm along the direction of the jet. The second simulation, which we denote the \spikes simulation, has a jet that was injected with similar initial conditions as the \steady jet with the exception of the jet being pulsed. The \spikes simulation jet was on for 40 half second pulses, each followed by another half second of quiescence. The luminosity of each pulse was decreased by 5\% with respect to the first pulse \citep{diego_lazzati_variable_grb}. The domain of the \spikes simulation is $2.56\times10^{12}$ cm along the jet axis. These simulations were previously analyzed by \cite{parsotan_mcrat} and \cite{parsotan_var} where they focus on various spectral properties and the synthetic light curves. In this paper we focus on the time integrated polarization and the time resolved polarization properties of each simulated GRB.

We ran the MCRaT code during the time in which the central engine of each model was active in order to investigate the effects of the varying central engine on the radiation. The number of photons injected into each simulation was $\sim 1.9 \times 10^7$ for the \steady simulation and $\sim 1.4 \times 10^6$ for the \spikes simulation. The order of magnitude difference was necessary to maintain a reasonable computation time for the \spikes simulation. This is a direct result of the jet in the \spikes simulation moving at $\Gamma \sim 10$ which is an order of magnitude smaller than the \steady simulation.

\begin{figure*}[t!]
\centering
\gridline{
\fig{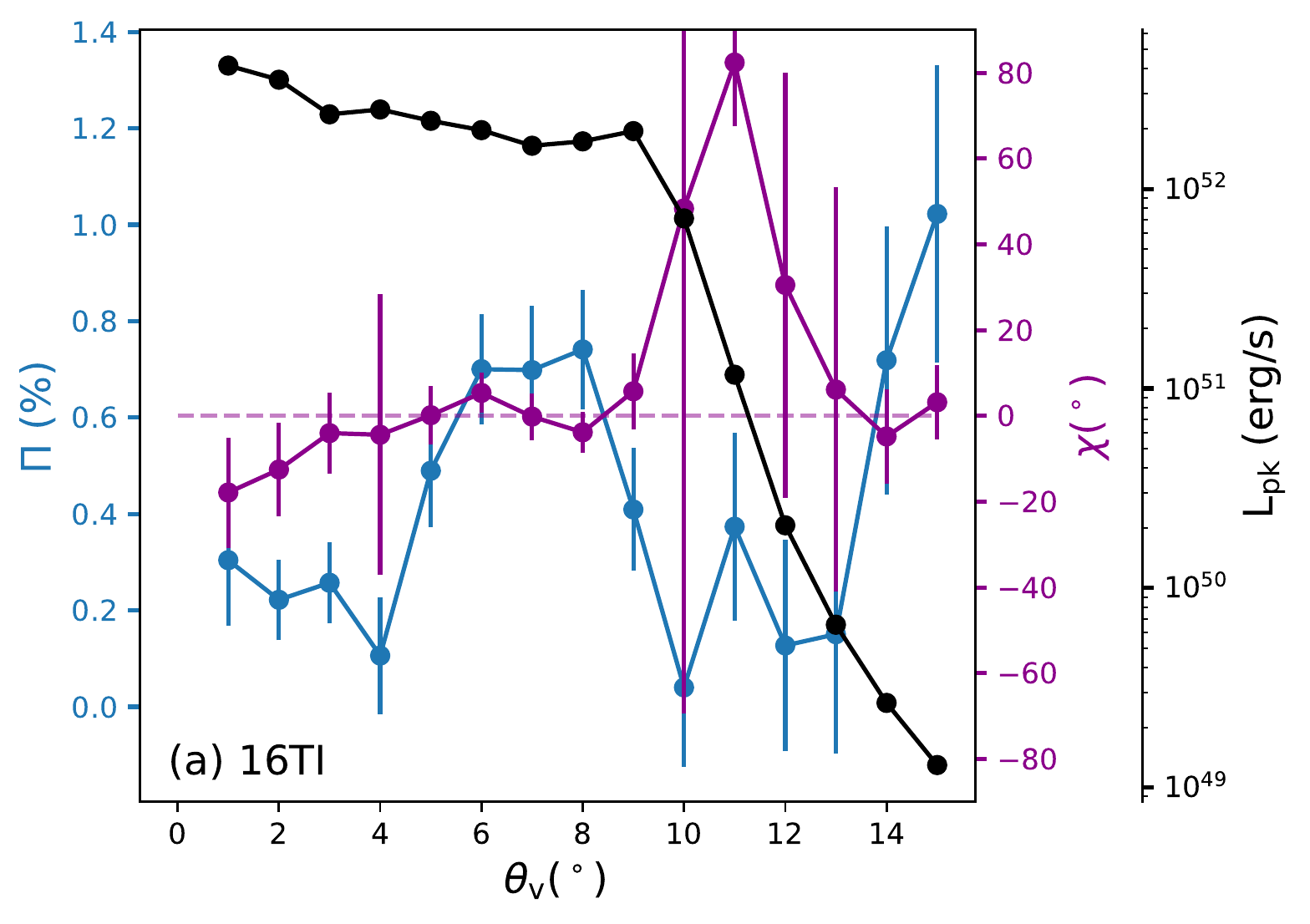}{0.5\textwidth}{\label{16ti_time_int_pol}}
\fig{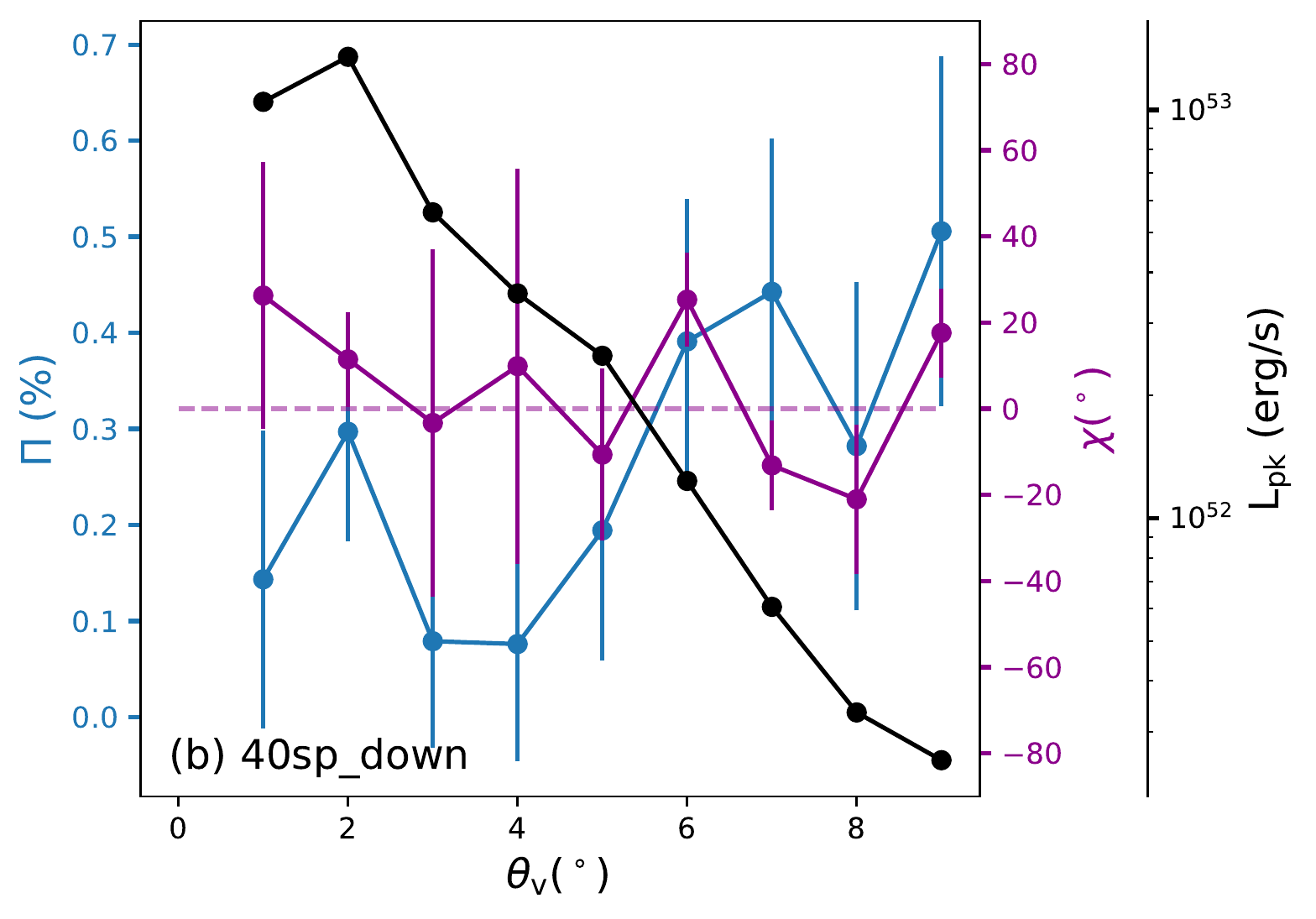}{0.5\textwidth}{\label{40sp_down_time_int_pol}}
}
\caption{Time integrated polarization degrees, plotted in blue, alongside the polarization angle, plotted in purple, and the peak {luminosity} of the light curve, {plotted in black,} at a given viewing angle, $\theta_{\mathrm{v}}$. The horizontal dashed line provides a reference for $\chi=0^\circ$. The left plot is for the \steady simulation and the right is for the \spikes simulation. We find that the time integrated $\Pi$ in both simulations is inversely proportional to $L_\mathrm{pk}$ of the synthetic GRB. Similar to Figure \ref{fig:complundmanp4thetaj1}, both figures shows a distinct decrease in $\Pi$ in the \steady simulation around $\theta_{\mathrm{v}} = 8^\circ$ which is due to the fact that the photons at these larger viewing {angles} are not fully decoupled from the flow. %This is not seen in Figure \ref{time_int_pol}(b) due to the fact that we do not inject photons at such large angles. 
%Accompanying this feature is a switch in the polarization angle being consistent with $0^\circ$ to then being consistent with $90^\circ$. 
}
\label{time_int_pol}
\end{figure*}

\begin{figure}[b]
\centering
\includegraphics[width=\linewidth]{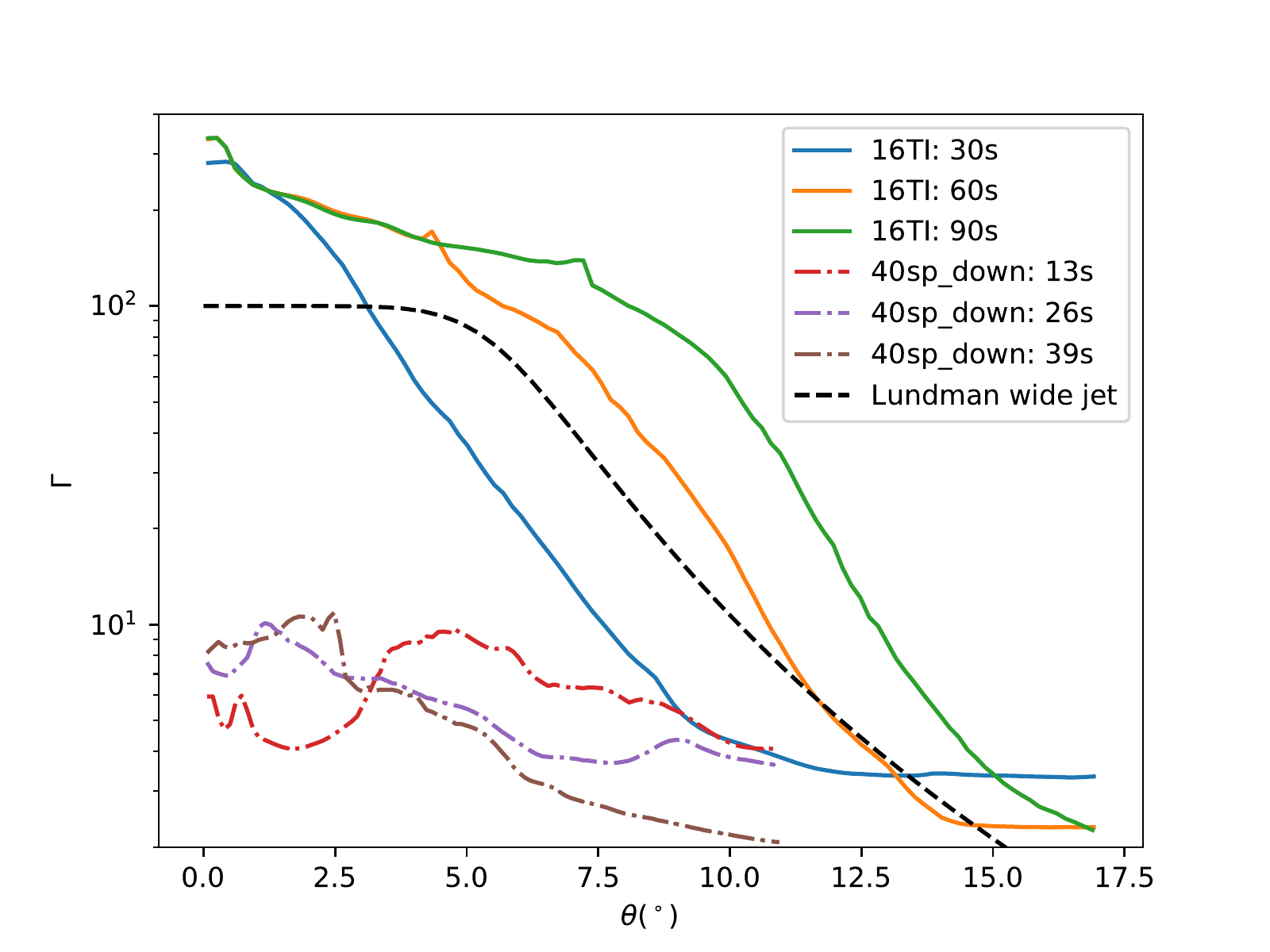}
\caption{Plot of the Lorentz factor, $\Gamma$, of the synthetic \steady and \spikes GRB jets as a function of angle at various times in its development. The times are associated with times that photons are detected in a given mock observed light curve. We also plot the $\Gamma$ profile of the wide jet as given by \cite{lundman2014polarization} with $\theta_{\mathrm{j}}=0.1$ and $\Gamma_0=100$. The \steady simulation is initially more compact than \citeauthor{lundman2014polarization}'s (\citeyear{lundman2014polarization}) jet but evolved to be wider, decreasing the detected $\Pi$. The \spikes simulation has a jet profile that is much wider than \citeauthor{lundman2014polarization}'s (\citeyear{lundman2014polarization}) wide jet. }
\label{fig:gammavstheta}
\end{figure}

\subsection{Time Integrated Polarization}

Figure \ref{time_int_pol} shows the time integrated polarization degrees, $\Pi$, angles, $\chi$, and {peak luminosity of the lightcurve}, $L_\mathrm{pk}$, as a function of observer viewing angles of the synthetic GRBs. {These results are acquired by integrating over the time that the jet is active.} Here, we plot $\Pi$ in blue, the light curve peak luminosity, $L_\mathrm{pk}$, in black, and $\chi$ in purple for the \steady simulation, in Figure \ref{time_int_pol}(a), and the \spikes simulation, in Figure \ref{time_int_pol}(b). $L_\mathrm{pk}$ is the peak of the synthetic light curve when the light curve is binned into 1 second time bins, analogous to the manner in which it is used in the Yonetoku relationship. In general, we find that $\Pi$ is negatively correlated to $L_\mathrm{pk}$, with spearman's rank coefficients $r_s=-0.6$ {\footnote{The $r_s$ values acquired are not statistically significant but they still provide a measure of how well correlated the variables are with respect to one another}} and $r_s=-0.27$ for the \steady and \spikes simulations, respectively. Additionally, $\Pi$ is positively correlated to $\theta_{\mathrm{v}}$, where  $r_s=0.65$ and $r_s=0.19$ for the \spikes and \steady simulation respectively. The direct relationship between $\Pi$ and $\theta_{\mathrm{v}}$ is easy to see in Figure \ref{time_int_pol}(b) for the \spikes simulation, however, it becomes more complicated in the \steady simulation. This is due to the fact that we see a turnover in the polarization in the \steady simulation at $\theta_{\mathrm{v}} = 8^\circ$, which is consistent with our verification results in Figure \ref{fig:complundmanp4thetaj1}. The photons at these larger angles are not fully decoupled from the flow, which means that their expected polarization degree is suppressed due to the ongoing scattering that is changing their Stokes parameters. {Excluding $\theta_{\mathrm{v}} > 8^\circ$ in the analysis of $r_s$ in the \steady simulation changes it to be $r_s=0.64$, which is consistent with the value acquired from the \spikes simulation.}
Accompanying this feature is a switch in the polarization angle being consistent with $0^\circ$ to then being consistent with $90^\circ$. 
%We also see this turnover in $\Pi$ in the \spikes simulation at $\theta_{\mathrm{v}} = 8^\circ$. 
 
The time integrated polarization is much smaller than what is found in other works. \cite{ito_polarization} and \cite{lundman2014polarization} find $\Pi$ as low as a few percent and as high as $\sim 40$ \%, for off axis observers. The difference between our study and theirs can be attributed to the structure of the jet in the RHD simulation and the fact that not all photons are reaching the photosphere (typically located at $\gtrsim 10^{13}$ cm); this is shown in Figures \ref{fig:complundmanp4thetaj1} and \ref{time_int_pol}. In Figure \ref{fig:gammavstheta} we show the structure of the \steady and \spikes jet as a function of angle at three different times in the jet's evolution, which correspond to various times in the synthetic light curves shown in the next section. The \steady simulation has a very fast core at all times. Initially, the jet is smaller than the wide jet presented by \cite{lundman2014polarization}, but the \steady jet eventually grows to become larger, which has the effect of decreasing the amount of polarization. We additionally find that the \spikes simulation has a relatively uniform lorentz factor profile as a function of angle, although it does vary in time, which is expected from a variable jet. Both of these synthetic GRBs have very wide jets that contribute to the extremely low time integrated polarization in Figure \ref{time_int_pol}; {these wide jets are relatively uniform and lacking steep gradients, a source of anisotropy in the flow, which means that there is little structure in the flow to produce very high time integrated polarization \citep{lundman2014polarization}. Furthermore, the variability in the \spikes simulation that should produce relatively large polarization ($\Pi\gtrsim1$\%) gets washed out as we integrate over the bright and dim portions of the light curve and what is left is the effect of the very wide jet profile that we observe in Figure \ref{fig:gammavstheta}.}

Although not all photons decouple from the jet, particularly at large $\theta_\mathrm{v}$ ($\gtrsim 9^\circ$), there is still enough of an anisotropy in the outflow to produce polarization that is significantly different from $\Pi \approx 0$\%. Since the anisotropies in the outflow will only increase as the jet becomes increasingly transparent to the radiation, the polarization is expected to increase. Thus, we consider all of the polarization measurements at large $\theta_\mathrm{v}$ in our results to be lower limits.

\subsection{Time Resolved Polarization}
\begin{figure*}[]
 \centering
 \gridline{
 \fig{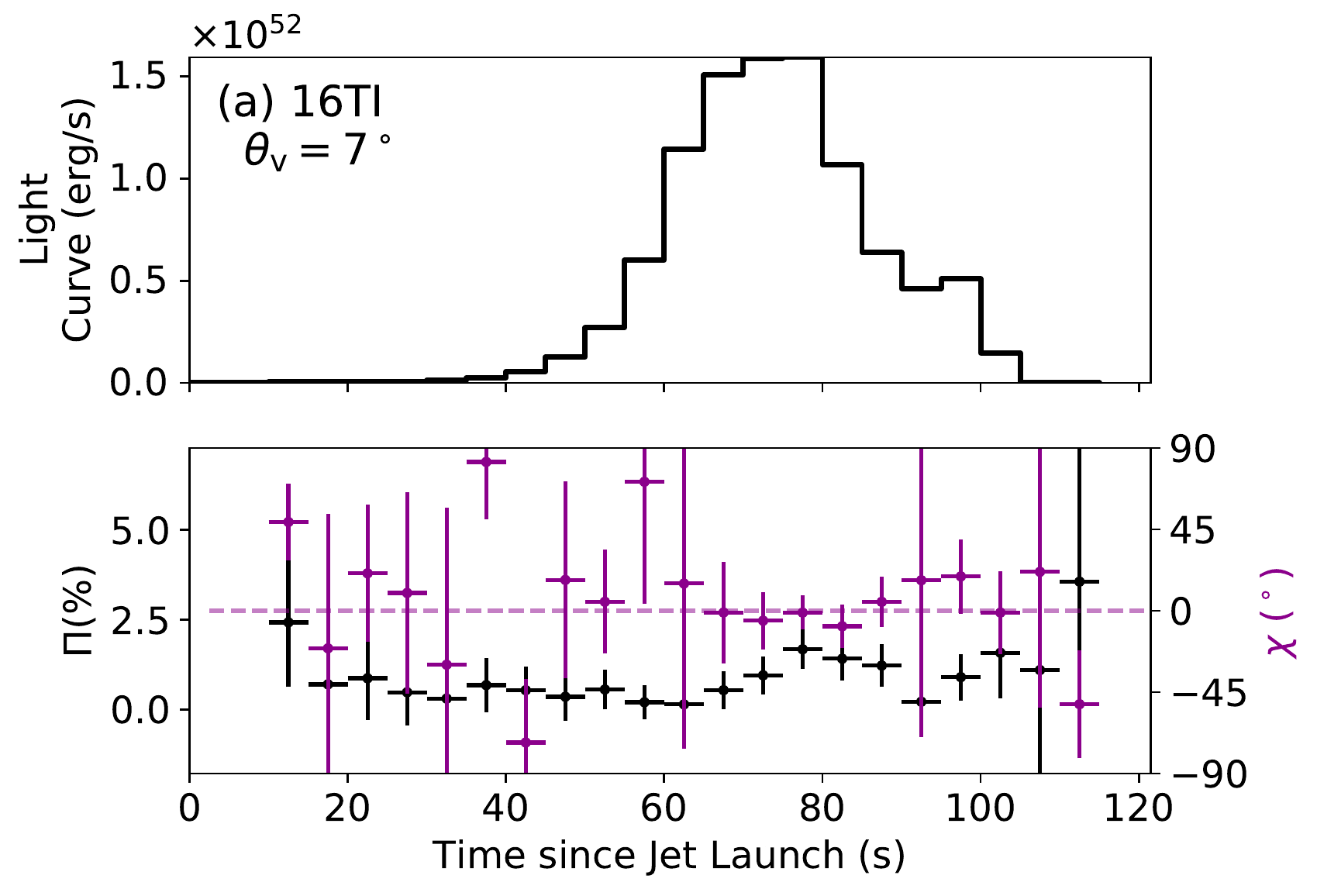}{0.5\textwidth}{\label{16ti_time_res_pol}}
 %}
 %\gridline{
 \fig{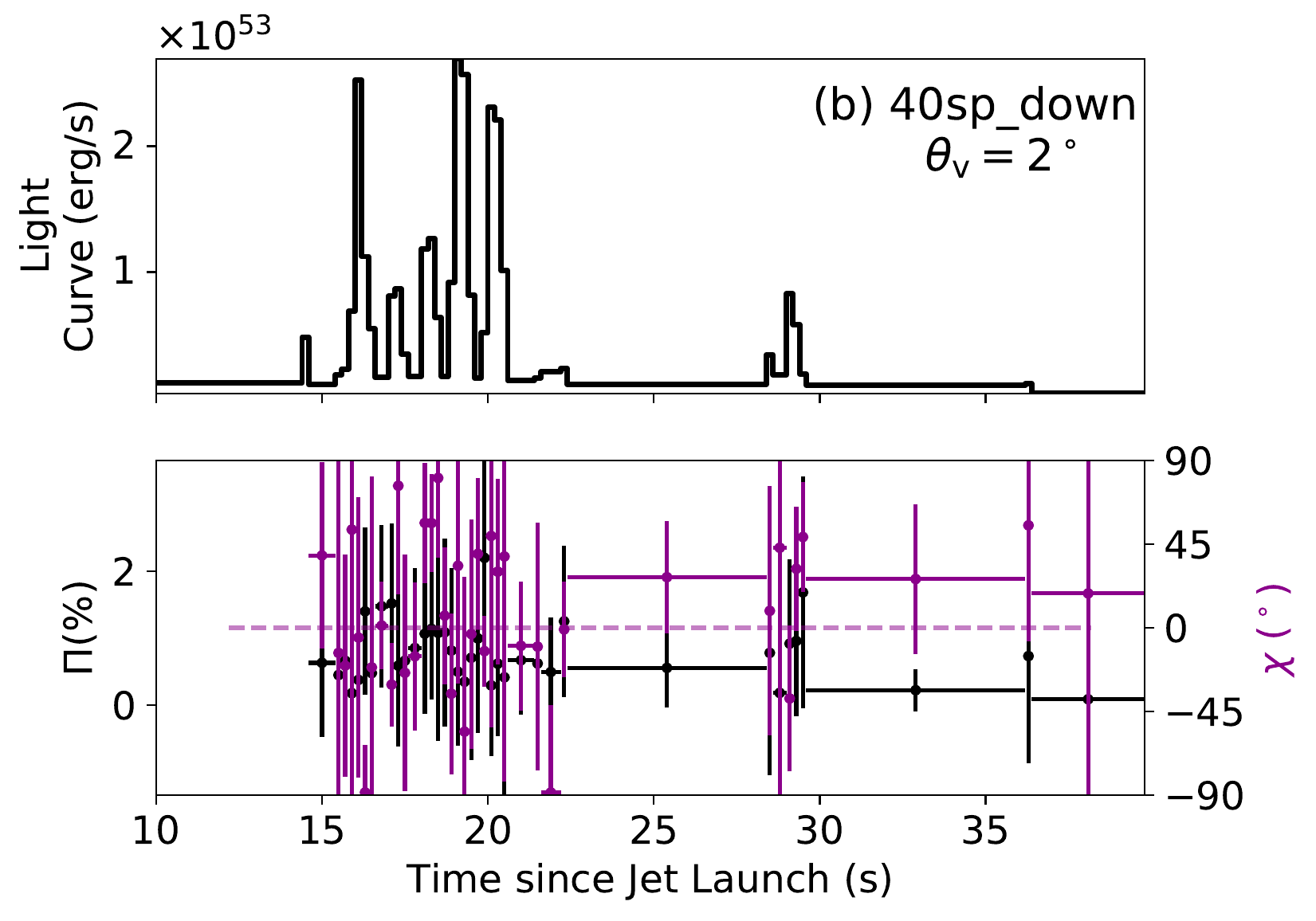}{0.5\textwidth}{\label{40sp_down_time_res_pol_2}}
 }
 \gridline{
  \fig{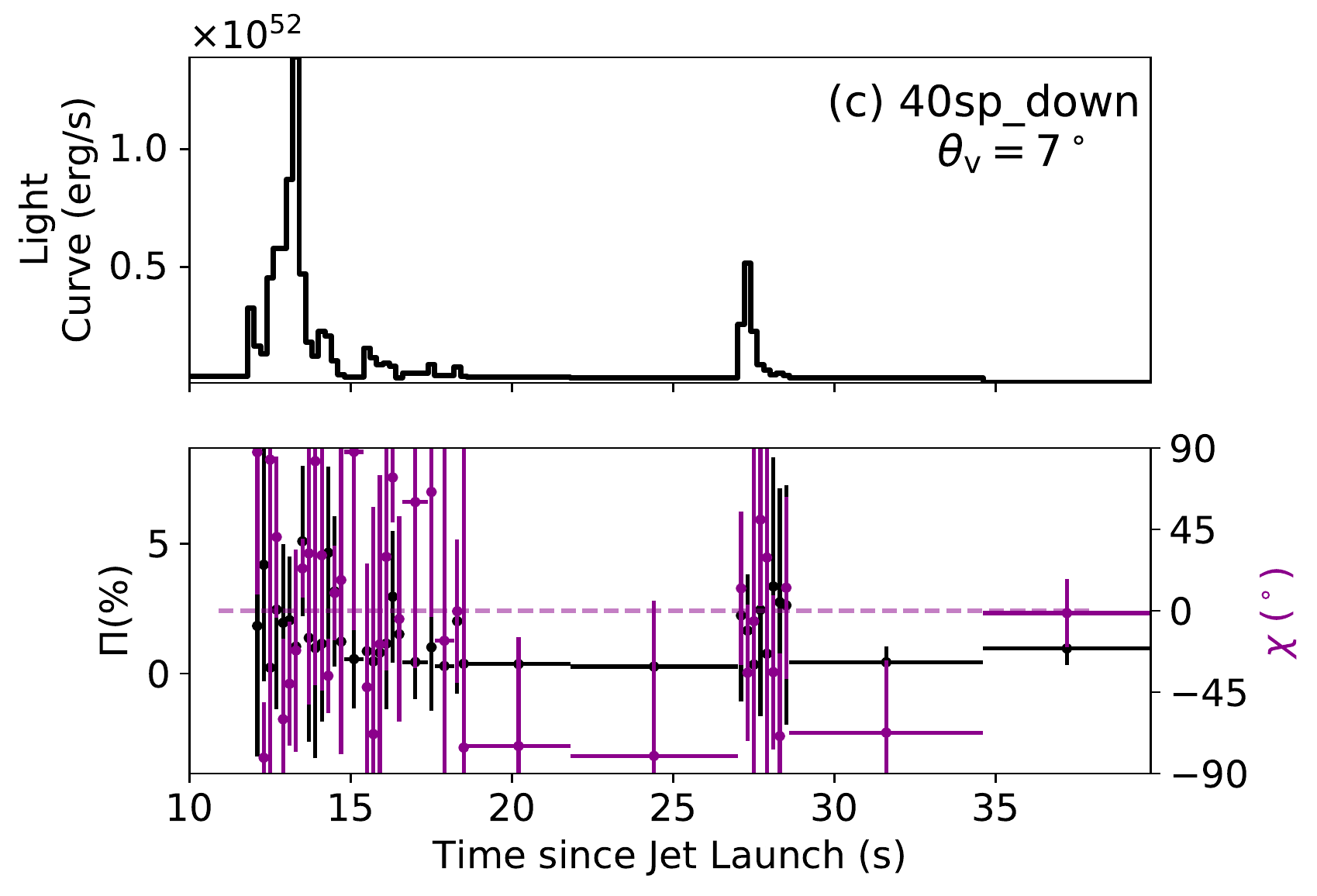}{0.5\textwidth}{\label{40sp_down_time_res_pol_7}}
    \fig{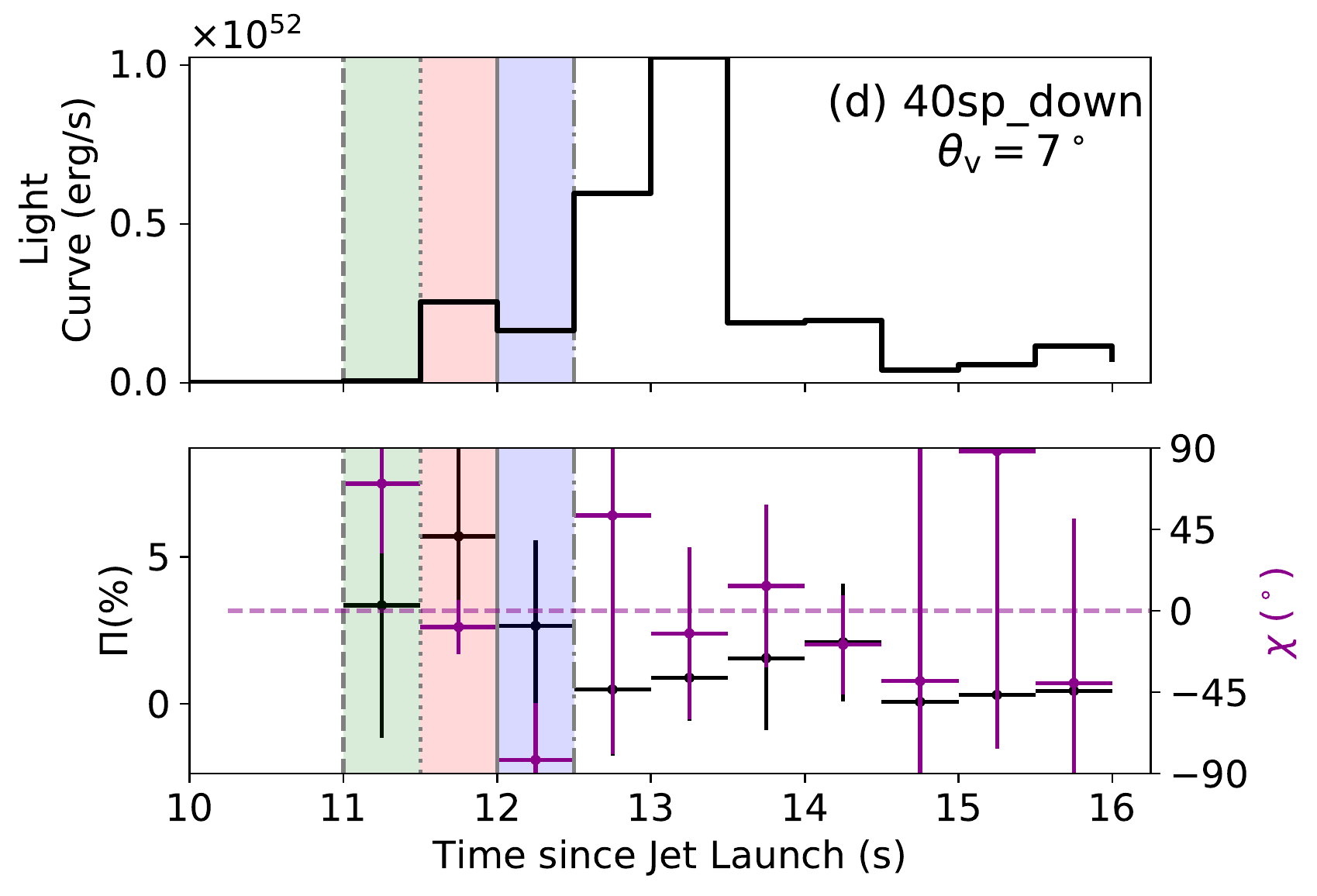}{0.5\textwidth}{\label{fig:40p_down_big_dt}}
  }
 \caption{Light curves, polarization degrees, in black in the bottom panel, and polarization angles, in purple, for the \steady simulation, Figure \ref{time_res_pol}(a), and the \spikes simulation for an observer located at $2^\circ$ and $7^\circ$, Figures \ref{time_res_pol}(b) and \ref{time_res_pol}(c) respectively. Figure \ref{time_res_pol}(d) shows Figure \ref{time_res_pol}(c) with a larger time bin of $0.5$ s. 
 %In the top panel the light curve is shown in black and the time resolved $E_\mathrm{pk}$ are shown in green, while in the bottom panel the polarization degree is shown in black and the polarization angle is shown in purple. 
 The horizontal dashed line in the bottom panels provides a reference for $\chi=0^\circ$. {The highlighted green, red and blue portions in Figure \ref{time_res_pol}(d), and the grey vertical lines, correspond to regions of equal arrival times shown in Figure \ref{time_surfaces} with similar lines.} The \steady simulation does not exhibit much polarization while the polarization angle stays around $0^\circ$ during the brightest portion of the light curve. The \spikes simulation show relatively high polarization degree at larger viewing angles in addition to evolution of $\chi$ within the synthetic GRB (Figure \ref{time_res_pol}(d) at $t\sim 11.5$ s). 
 %During the brightest period of the \spikes GRB, similar to the \steady simulation, $\chi$ stays approximately constant. 
 We find that a changing $\chi$ is indicative of various shells of material coming into the line of sight of the observer due to structure within the jet, while a constant $\chi$ indicates that the emission region is staying relatively constant due to lack of structures within the jet.} 
 \label{time_res_pol}
 \end{figure*}
 
 \begin{figure*}[t!]
\centering
\includegraphics[width=\linewidth]{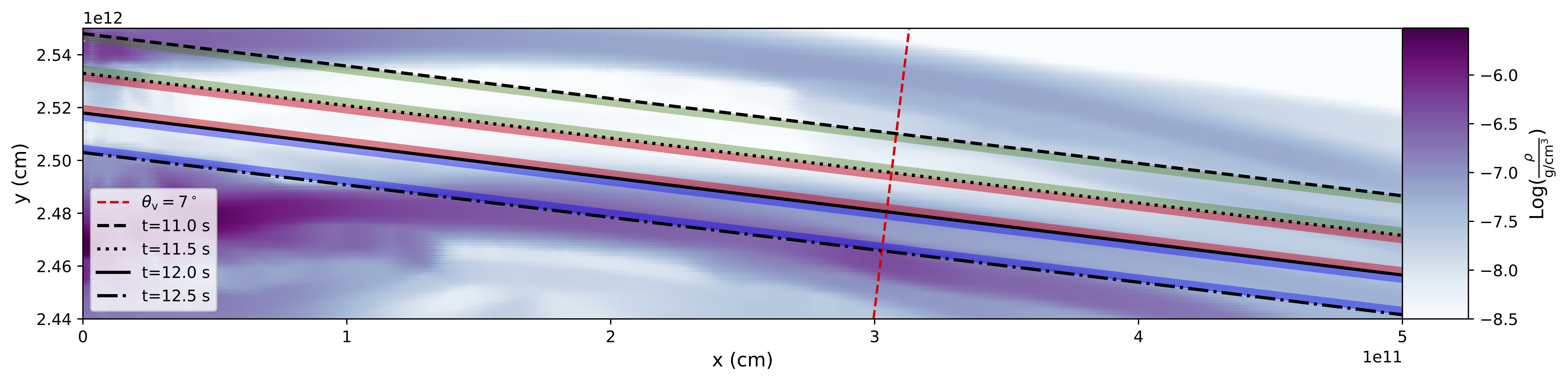}
\caption{A pseudo-color density plot of a region of the \spikes GRB simulation. The red dotted line corresponds to the line of sight of an observer located at $\theta_\mathrm{v}=7^\circ$. The various  lines are surfaces of equal arrival times that are detected by an observer at $\theta_\mathrm{v}=7^\circ$ at various times. {The highlighted regions, colored green, red, and blue, correspond to the start and end of the same colored regions shown in Figure \ref{time_res_pol}(d).} We find that the change in $\chi$ seen in Figure \ref{time_res_pol}(d) is due to seeing different portions of the jet at various times. Initially, at $t=11$ s, the observer sees the core of the jet. Then they observe photons originating from outer regions of the jet at $t=11.5-12$ s. Finally, the observer sees more of the inner region of the jet again by $t=12.5$ s.
}
\label{time_surfaces}
\end{figure*}
 
An advantage of using MCRaT on a time dependent synthetic 2D RHD GRB  jet is the ability to produce time resolved polarization predictions. Figure \ref{time_res_pol} shows the time resolved polarizations of the \steady simulation and the \spikes simulations at two different observer angles. In the top panel we plot the light curves in black. In the bottom panel we show the time resolved polarization degree in black and the time resolved polarization angle in purple. The dotted purple line is the $\chi = 0^\circ$ reference line. The \steady simulation is binned uniformly in time with $dt=0.5$ s and the \spikes results are binned non-uniformly with a binning criteria based on the total energy received in each time bin. This energy must be larger than some critical luminosity before the end of the time bin is determined. Figure \ref{time_res_pol}(b) and \ref{time_res_pol}(c) have luminosity cutoffs of $2 \times 10^{52}$ erg/s and $2 \times 10^{51}$ergs/s, respectively.

For the case of the \steady GRB simulation, shown in Figure \ref{time_res_pol}(a), the polarization is very small at all times ($\Pi \lesssim 1.5$ \%). This can be attributed to the lack of structure in the jet, as is shown in Figure \ref{fig:gammavstheta}. The brighter portion of the light curve is indicative of low optical depth regions of the outflow \citep{parsotan_var}; it is during this period that we get the most well constrained polarization measurements (see Section \ref{methods} where the errors are $\propto N^{-\frac{1}{2}}$). We see that $\Pi$ is also at its maximum, at $\sim 2$\%, and $\chi \approx 0^\circ$ at all times during the maximum of the light curve. The \steady simulation does not have multiple shells of material coming into view of the observers line of sight. As a result, we can use it as a control for analyzing the \spikes simulation.

{Things are much different in the case of \spikes, where the jet is variable and there is lots of time dependent spatial structure} on the scale of $\delta \theta \sim  \Gamma^{-1}$, where $\Gamma$ is $\sim 10$ (see Figure \ref{fig:gammavstheta}). At $\theta_{\mathrm{v}}=2^\circ$, in Figure \ref{time_res_pol}(b), we see that $\Pi$ is relatively small with a maximum of $\sim$2\%. At this viewing angle we are in the core of the jet where the profile is relatively symmetric and, as a result, do not have much interference with off axis shells of material coming into the observers line of sight. This case is very similar to the \steady simulation analyzed in Figure \ref{time_res_pol}(a).

Figure \ref{time_res_pol}(c) shows the \spikes light curve, $E_\mathrm{pk}$, $\Pi$, and $\chi$ for an observer at $\theta_{\mathrm{v}}=7^\circ$. In this case, we observe larger $\Pi$, with the maximum being $\sim 5$\%. Additionally, the polarization angle changes multiple times during this synthetic GRB observation. On order to emphasize this oscillation in $\chi$,  we replot the largest pulse in Figure \ref{time_res_pol}(c) using larger time bins of $dt=0.5$ s, in Figure \ref{time_res_pol}(d). We see that within the first few seconds of the GRB, $\chi$ changes from $90^\circ$ to $0^\circ$ and back to $90^\circ$ within a $\delta t \sim 1$ s. {Then, $\chi \approx 0^\circ$ during the brightest portion of the light curve, at $t\approx13$ s.} The constant value of $\chi$ is similar to what we see for the \steady simulation where there is little structure in the jet and we are seeing the main emission region along the observer's line of sight (LOS). On the other hand, the changing $\chi$ suggests that other shells of material within the jet are coming into the observer's line of sight due to variability in the jet. 

In order to confirm that a changing $\chi$ is representative of various regions of the jet coming to the observers LOS, we relate the times in the light curve shown in Figure \ref{time_res_pol}(d) to equal arrival time surfaces within the jet. In Figure \ref{time_surfaces} we plot the density in a region of the jet that corresponds to the regions that are emitting during the times of interest in the light curve. These equal time of arrival surfaces are shown as a variety of different colored highlighted regions that correspond to the vertical highlighted regions plotted in Figure \ref{time_res_pol}(d). {We find that at $t=11.0-11.5$ s the dense jet core {($\rho \sim 10^{-6}$ g/cm$^3$ at $y\sim 2.54\times 10^{12}$ cm and $x\sim 0$ cm)} is the dominating emission material, scattering photons into the observers LOS \citep{parsotan_var} with $\chi \approx 90^\circ$. At $t=11.5-12$ s, the emitting region is between two dense shells in the jet and the observed photons are primarily originating from the denser outer region of the jet, where $\rho \sim 10^{-7.5}$ g/cm$^3$ located at $y\sim 2.47\times 10^{12}$ cm and $x\sim 5\times 10^{11}$ cm, changing $\chi$ to be $ \approx 0^\circ$. Finally, by $t=12.5$ s, the observer sees emission from the inner region of the jet again, near $x\sim 2 \times 10^{11}$ cm and $y \sim 2.48\times 10^{12}$ cm where $\rho \sim 10^{-6.5}$ g/cm$^3$, thus changing the polarization angle once more.} The width between the two dense shells of material corresponds to the $c \delta t$ of the changing $\chi$. The outlined effect of a changing $\chi$ as related to the lateral and temporal structure of the jet is seen by an observer located at $\theta_\mathrm{v} = 3^\circ-8^\circ$. The upper limit is related to the fact that photons aren't fully decoupled from the flow at $\theta_\mathrm{v} \gtrsim 9^\circ$ and the lower limit is due to the core of the jet being the dominating emission region at all times for $\theta_\mathrm{v} \lesssim 3^\circ$.

{The changing $\chi$ that we observe are consistent with the observed changing polarization angle of GRB 170114A \citep{zhang2019polar}. These changing $\chi$ from the \spikes simulation also suggests that the jet that produced GRB 170114A was variable. }

%The $\delta t \sim 1$ s implies that the shell width is $\delta r \sim 3\times 10^{10}$ cm. When we look at the \steady RHD simulation, we find a shell of denser material, with a similar $\delta r$, that would be emitting photons at $t \sim 12$ s in the light curve. This denser shell allows photons from other regions of the jet to be brought into the observer's line of sight \citep{parsotan_var}, changing $\chi$. The front portion of the shell that emits photons first has $\chi=90^\circ$, then in the middle of the shell $\chi=0^\circ$, and finally at the back edge of the shell $\chi=90^\circ$ again. As a result, the cross correlation between the FLASH simulation and the radiative transfer light curve shows that the time that it takes $chi$ to change can be a proxy for the width of shells found in a variable GRB jet. 

%We see this effect at other $\theta_{\mathrm{v}}$ between $3^\circ$ and $7^\circ$. At $\theta_{\mathrm{v}}<3^\circ$ the shell doesnt exhibit any gradients

\section{Summary and Discussion} \label{summary}
We have used the MCRaT code to conduct Monte Carlo radiation transfer simulations of time resolved and time integrated polarization in LGRBs. MCRaT injects, propagates, and compton scatters photons using the full Klein Nishina cross section. These photons are injected into an outflow described by a FLASH 2D RHD simulation and are subsequently scattered until the end of the simulation. This process of injecting, propagating and scattering photons is repeated until there are no more photons to be injected in the simulation. The implementation of polarization in MCRaT was verified in multiple ways and we were able to recover the time integrated polarization profile of a wide jet as presented by \cite{lundman2014polarization}. We ran the MCRaT simulations using the RHD simulations of a steady GRB jet, the \steady simulation, and a variable engine jet, which we denote as the \spikes simulation.

Primarily, we have found that:
\begin{itemize}
\item Not all photons at large $\theta_\mathrm{v}$ ($\gtrsim 9^\circ$) in the MCRaT simulation are decoupled from the synthetic GRB jet which decreases the mock observed $\Pi$; as a result, the $\Pi$ presented in this paper at large $\theta_\mathrm{v}$ are lower limits.
\item The time integrated polarizations are generally positively correlated with $\theta_{\mathrm{v}}$, {with $r_s\approx 0.65$ for both simulations,} and negatively correlated with $L_\mathrm{pk}$, with the correlation between $\Pi$ and $L_\mathrm{pk}$ being $r_s=-0.6$ and $r_s=-0.27$ for the \steady and \spikes simulations respectively.
\item The \steady simulation has very little structure on the scale of $\Gamma^{-1}$ which decreases $\Pi$ and contributes to $\chi$ being approximately constant in time.
\item  The \spikes simulation shows more structure temporally and spatially which contributes to various shell of materials coming into the line of sight of the observer, as a result, this simulation has larger time resolved $\Pi \sim 5\%$ and a changing $\chi$.
\item The $\delta t$ for $\chi$ to change in a variable jet is indicative of the width of shells of materials within the GRB jet.
\item A changing $\chi$ also indicates that the observer is seeing various regions of the GRB jet, which may help constrain temporal and lateral variability in the jet structure.
\end{itemize}

The MCRaT results of the time integrated quantities {($\Pi \lesssim 1$\% and $\chi \approx0^\circ$)} are consistent with the results presented by \cite{lundman2014polarization}{, where wider jets produce lower polarization degrees}, however our results show the importance of ensuring that all photons are decoupled from the flow. Our simulations have a finite domain which prevents some photons at $\theta_\mathrm{v} \gtrsim 9^\circ$ from being decoupled from the flow by the time they propagate to the edge of the simulation domain. As \cite{parsotan_var} mentioned, the peaks in the light curves are composed of photons that are mostly decoupled from the outflow while the quiescent portions are still coupled to the flow. The dimmer regions contribute to lowering the overall time integrated $\Pi$, although we do detect some polarization from these dimmer times in the light curves due to structure in the GRB outflow \citep{lundman2014polarization}. As the photons become less coupled to the outflow, the asymmetry in the jet will become more pronounced with respect to the radiation and we will expect to detect larger $\Pi$. This drives the need to conduct larger domain RHD simulations, to ensure that the photons at larger $\theta_\mathrm{v} (\gtrsim 9^\circ)$ are decoupled from the flow and we have an accurate mock observation of $\Pi$ at these large angles.

The results presented here are consistent with the results found by \cite{zhang2019polar}. Our mock detected $\Pi$ are within the limits that they find for their GRBs ($\Pi \lesssim 10$ \%). {In particular, \cite{zhang2019polar} provide a 99$^\mathrm{th}$ percentile upper limit of $\Pi<28\%$ for GRB 170114A, but quote $\Pi \approx 4$\% which is consistent with the $\Pi$ that we acquire in our study.} The low $\Pi$ that we find is also consistent with the findings of \cite{Iyyani_low_pol} for the first time period of GRB 160325A.

Additionally, we are able to show that variable GRB jets can produce changing $\chi$ such as is observed in GRB 170114A. \cite{zhang2019polar} report the $\chi$ of the aforementioned GRB to be $122^\circ$ for the first portion of the burst and $17^\circ$ for the second portion. While our polarization angles can only change between $0^\circ$ and $90^\circ$ due to the symmetry of the simulations, we may see a continuous change in $\chi$ from 3D MCRaT simulations.
%, although this remains to be seen theoretically. 
Additionally, \cite{zhang2019polar} do not specify any errors on their values of $\chi$ in GRB 170114A, however, their plots of the fitted $\Pi$ and $\chi$ parameters' confidence contours show that $\chi$ is not well constrained. As a result, it is still possible that $\chi$ is $\sim 0^\circ$ or $\sim \pm 90^\circ$ which is expected for an axis-symmetric jet, such as the ones studies in this work. {The fact that the \spikes simulation produces a changing $\chi$ suggests that GRB 170114A had a variable jet, with various parts of the jet coming into the observer's line of sight.} Our results are also able to explain the changing $\chi$ seen by \cite{Iyyani_changing_chi} for GRB 160821A.

One important assumption that \cite{zhang2019polar} make is that $\Pi$ is constant throughout a GRB and $\chi$ can change in time. In our analysis we find that this assumption may not be valid (see for example Figure \ref{time_res_pol}(d) where both $\Pi$ and $\chi$ change). Even if the GRB has a steady jet being injected (which is not known a priori), $\chi$ will be approximately constant and $\Pi$ may change (see Figure \ref{time_res_pol}(a)).

{
With future GRB polarimetry missions, such as POLAR-2 \citep{kole2019polar} and LEAP \citep{mcconnell2017leap_polarimeter}, expected to observe GRB polarizations with higher precision than that of the POLAR detector, we can use our findings to place constraints on the expected $\Pi$ and $\chi$ of the photospheric model compared to other GRB emission models. Globally, we expect the polarization angle, $\chi$, to be $\sim 0^\circ$ or $\sim 90^\circ$. While it is possible that the photospheric model may be able to account for other values of $\chi$ in 3D, it is unlikely that the polarization angle will deviate significantly from being aligned perpendicular or parallel to the plane defined by the jet axis and the observer's line of sight. Thus, a measurement of $\chi$ that is significantly different from these values combined with a large value of $\Pi$ may indicate that there is a global magnetic field in the jet \citep{toma2008statistical_GRB_pol}. In the case of a random magnetic field, the observed $\Pi$ will be low and the observed value of $\chi$ would be $\sim 0^\circ$ \citep{toma2008statistical_GRB_pol, Gill_Polarization}, similar to the photospheric case. The ability of the photospheric model to account for very low and high values of time integrated polarization degrees based on the geometry of the jet \citep{lundman2014polarization}, makes it very difficult to distinguish from the synchrotron model, with a global or random magnetic field, based solely on that observed parameter. As a result, time resolved $\Pi$ and $\chi$ observations becomes much more important. \added{The results acquired in this work suggests that GRBs observed on axis will have small $\Pi \lesssim 2\%$ and constant $\chi$, during the brightest portion of the light curve, while GRBs observed off axis will have larger $\Pi$ and $\chi$ that change by $\sim 90^\circ$ with the temporal structure of the GRB.} If the emission mechanism is synchrotron with a global magnetic field, the $\Pi$ and $\chi$ should not change in time; however with a random magnetic field, $\Pi$ and $\chi$ may vary randomly based on the magnetic field configuration of the GRB at a given time at a given observer line of sight \citep{Gill_Polarization}. There still needs to be more research conducted to produce robust time resolved polarization degree and angle predictions for a variety of jet structures and magnetic field configurations, for each emission model.
%For robust constraints to be placed on the radiation mechanism from time domain observations of these parameters, there needs to be additional predictions of the time resolved $\Pi$ and $\chi$ signatures of synchrotron emission from global magnetic field and a random magnetic field 
}

Unlike the analysis conducted by \cite{lundman2018polarization}, we do not consider the polarization at various energy bands. \cite{lundman2018polarization} has shown that this can be an interesting area of testing the photospheric model and the capability of testing time resolved polarization of different energy bands can also prove to be fruitful. With this goal in mind, and to ensure that there are enough low energy photons in the outflow \citep{parsotan_mcrat, parsotan_var},  MCRaT will be modified to consider the effects of synchrotron radiation and absorption. This improvement, combined with larger domain RHD simulations will allow us to make stringent time integrated and time resolved predictions of GRB spectra and polarizations.

\acknowledgements We would like to thank Hirotaka Ito, Christoffer Lundman, and Henric Krawczynski for discussion about the Stokes parameters and ways to ensure that MCRaT properly accounts for the Stokes parameters in every frame of reference. We thank them for sharing their codes with us. 

TP and DL acknowledge support by NASA grants 80NSSC18K1729 (Fermi) and NNX17AK42G (ATP), Chandra grant TM9-20002X, and NSF grant AST-1907955. TP acknowledges funding from the Future Investigators in NASA Earth and Space Science and Technology (FINESST) Fellowship, NASA grant 80NSSC19K1610. Resources supporting this work were provided by the NASA High-End Computing (HEC) Program through the NASA Advanced Supercomputing (NAS) Division at Ames Research Center. Additionally, this work used the CoSINe High Performance Computing cluster,  which is supported by the College of Science at Oregon State University. D.L.C is supported by C{\'a}tedras CONACyT at the Instituto de Astronom{\'i}a (UNAM) and acknowledges the support from the Miztli-UNAM supercomputer (project LANCAD-UNAM-DGTIC-321). %\newline

\bibliography{references}
%\newpage
\appendix
\restartappendixnumbering
\section{Comparing MCRaT to Other Analysis}
\label{comp_mcrat}
To show that the cross section sampling algorithm and Lorentz boosting of the Stokes parameters are correct, we reproduce some results acquired by \cite{depaola2003new} and \cite{krawczynski2011polarization}. 

\begin{figure}[h!]
 \centering
 \includegraphics[]{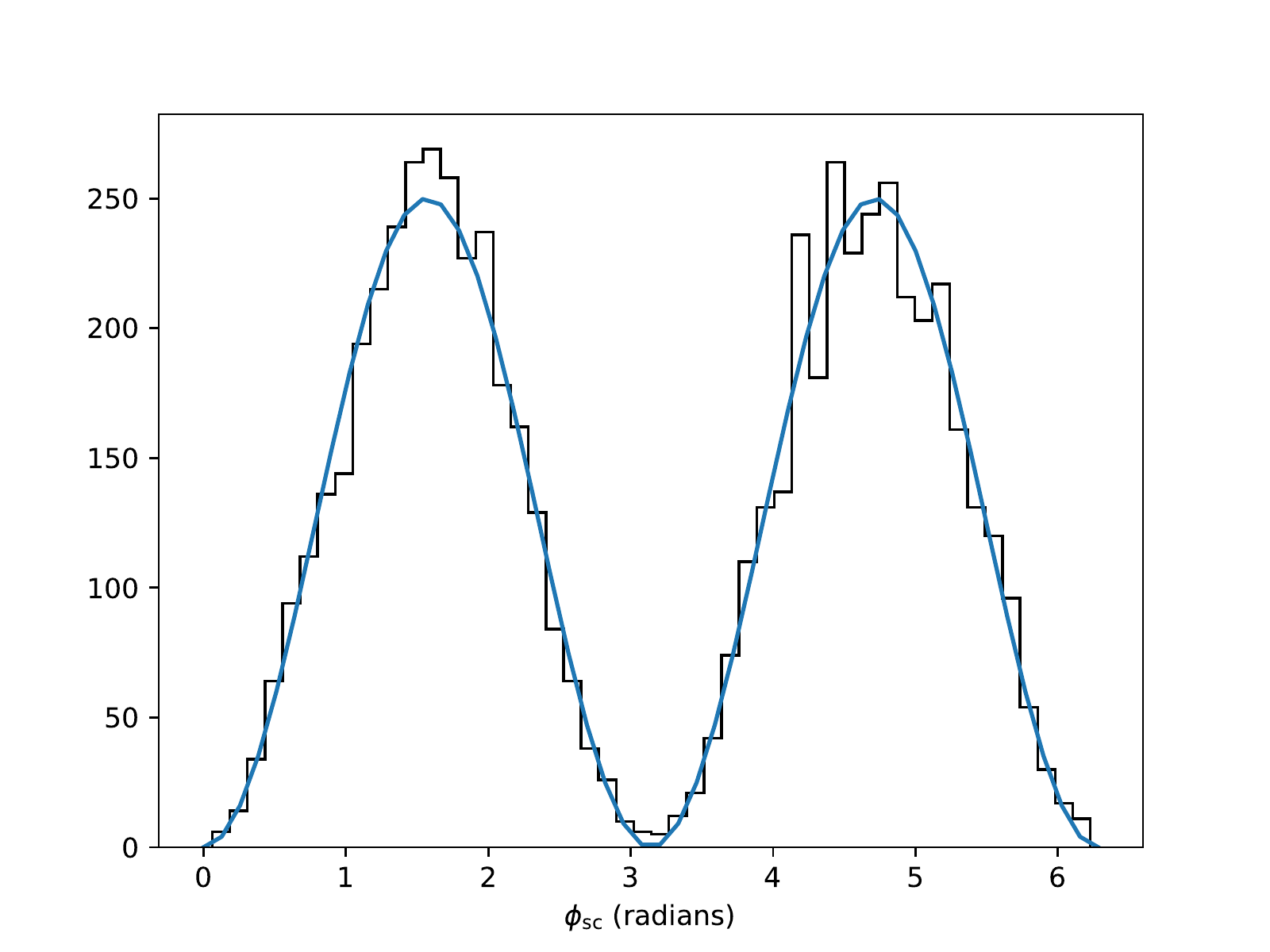}
 \caption{Plot of the distributed $\phi_\mathrm{sc}$ from sampling the Klein Nishina cross section as is outlined in Section \ref{mcrat} compared to the analytic cross section acquired by \cite{depaola2003new}. The distribution is shown in black and the analytic profile is shown by the blue curve.}
 \label{depaola_comparison}
 \end{figure}

Figure  \ref{depaola_comparison} shows the resulting modulation curve when our Klein Nishina (KN) cross section is monte carlo sampled in black. The analytic profile of the cross section as a function of $\phi_\mathrm{sc}$ calculated by \cite{depaola2003new} is shown as the solid blue line. They sample the KN cross section for a photon beam with 100\% polarization along the $+Q$ direction, with an energy of 100 keV and $85^\circ<\theta_\mathrm{sc}<90^\circ$. Accounting for the differences in the positive stokes parameters in the convention used, we are able to use our method of sampling the stokes parameter to acquire the proper distribution.

\begin{figure}[]
 \centering
 \gridline{
 \fig{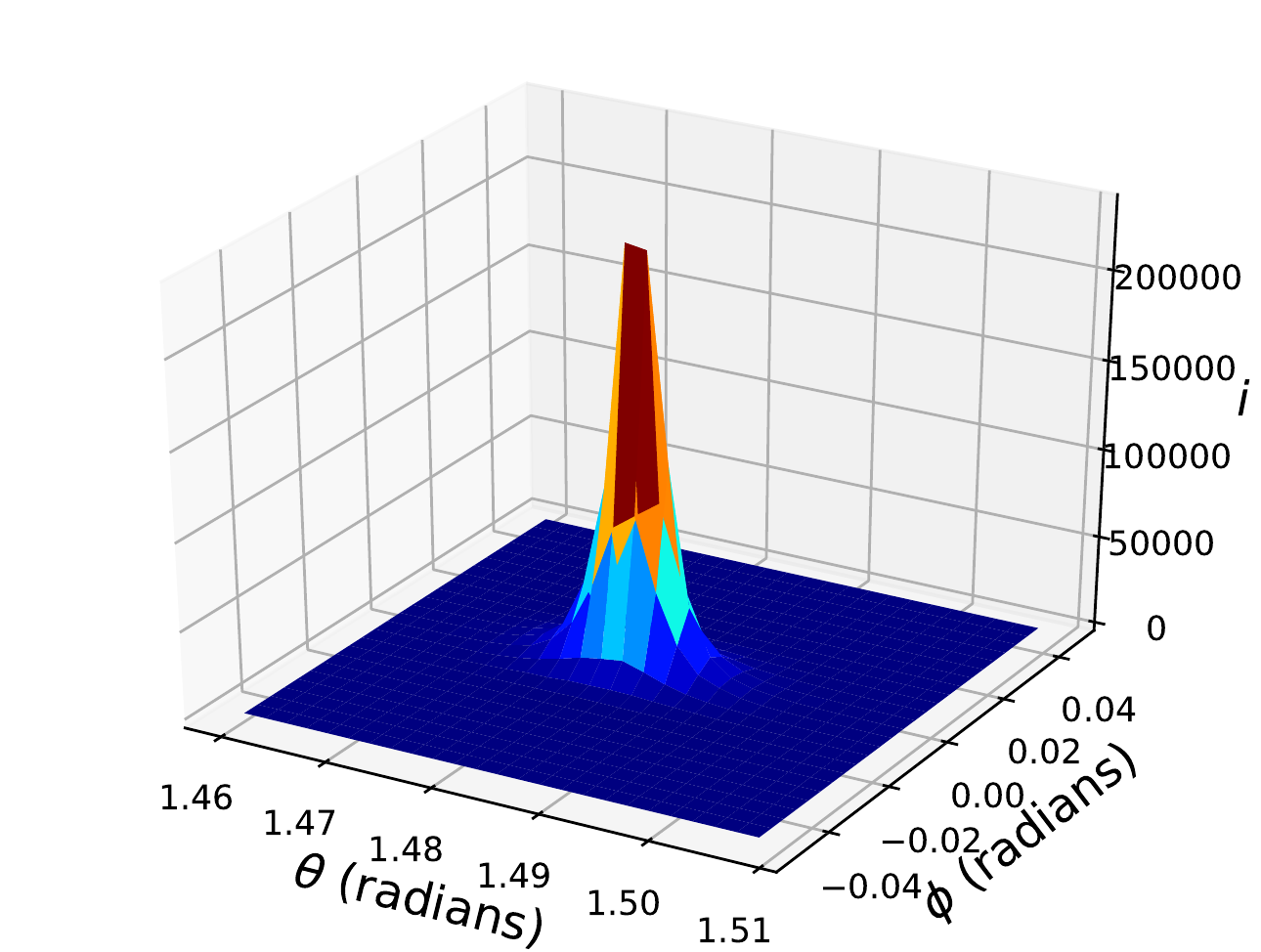}{0.5\textwidth}{\label{kraw_i}}
 %}
 %\gridline{
 \fig{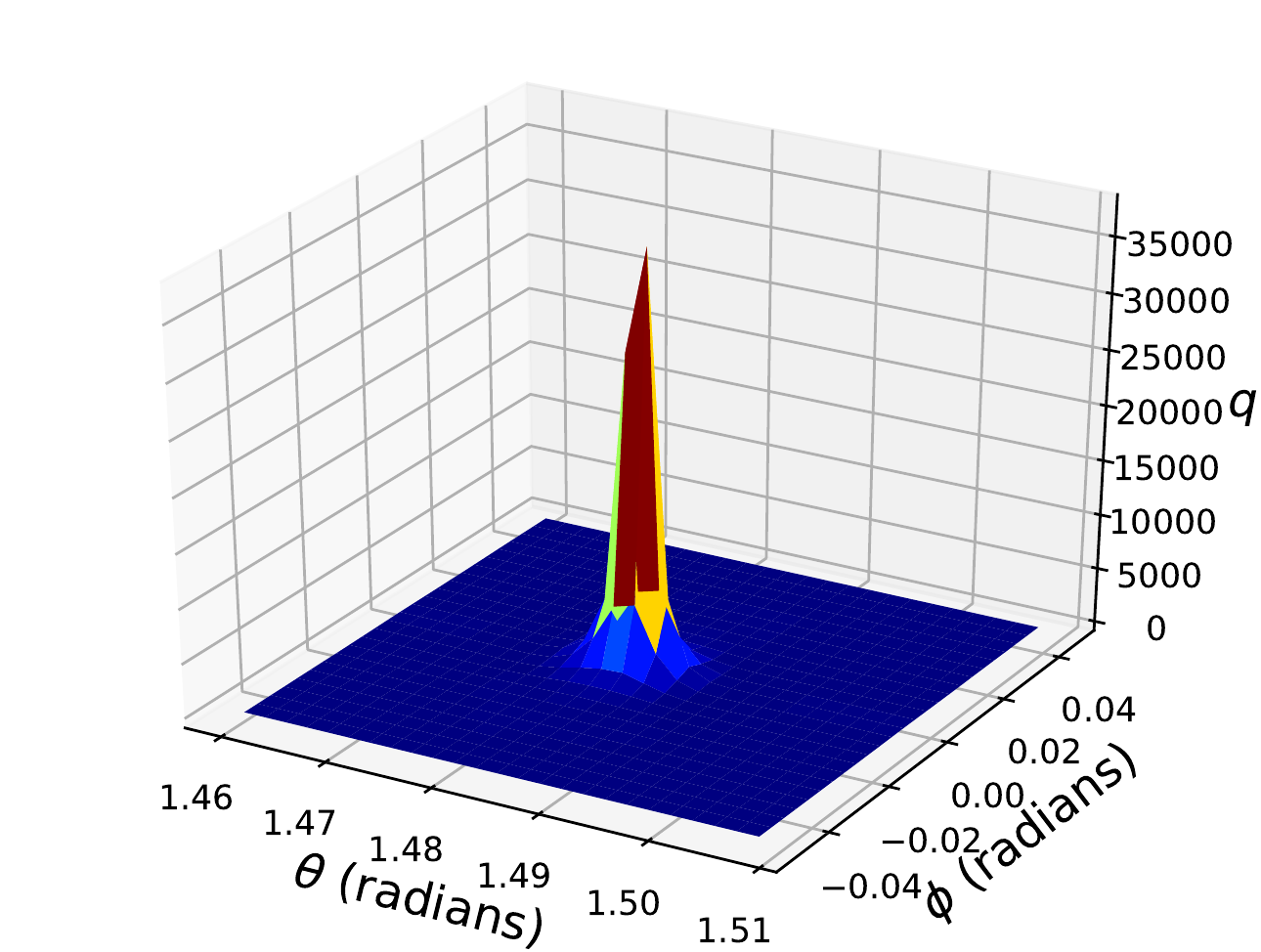}{0.5\textwidth}{\label{kraw_q}}
 }
 \gridline{
  \fig{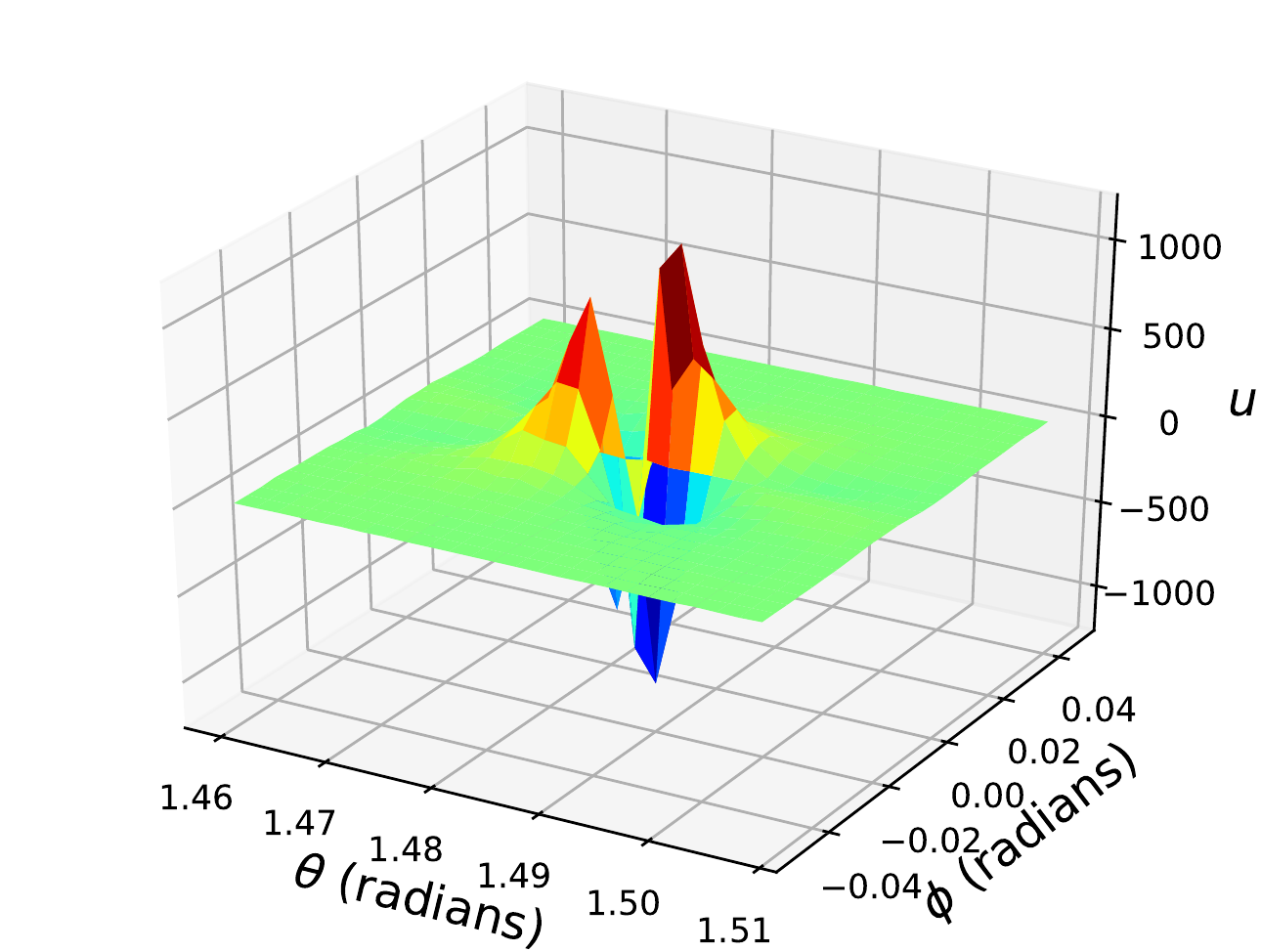}{0.5\textwidth}{\label{kraw_u}}
  }
 \caption{The distribution of the stokes parameters as a function of the scattered photon's $\theta$ and $\phi$ values. These distributions are morphologically similar to the distribution acquired by \cite{krawczynski2011polarization} in thier Figure 6.  }
 \label{kraw_comparison}
 \end{figure}

In order to verify the algorithm of switching frames of reference, we reproduce the results acquired by \cite{krawczynski2011polarization}. \cite{krawczynski2011polarization} scattered a beam of photons, 100\% polarized along the $+Q$ axis with frequency $\omega=10^{12}$ Hz, with a beam of electrons moving at a Lorentz factor of $\gamma=100$. Following their setup, we produce distributions of the resulting lab frame stokes parameters as a function of scattering angles, which we show in Figure \ref{kraw_comparison}. These distributions are identical, with the exception of normalization, to the distribution acquired by \cite{krawczynski2011polarization} in their Figure 6. The difference in the signs of $Q$ and $U$ in our distribution with respect to \citeauthor{krawczynski2011polarization}'s (\citeyear{krawczynski2011polarization}) Figure 6 is due to the differences in the orientation of the $+Q$ and $+U$ axis of the Stokes plane.

 \listofchanges
\end{document}